\documentclass[twoside]{article}
\usepackage{fleqn,espcrc2}

\usepackage{graphicx}

\usepackage{latexsym}

\newcommand{\AmS}{{\protect\the\textfont2
  A\kern-.1667em\lower.5ex\hbox{M}\kern-.125emS}}

\title{Domain wall fermion calculation of nucleon 
\(g_{_A}/g_{_V}\)\thanks{Presented by
S. Ohta at {\it Lattice 2000\/}, Bangalore, India, for
RIKEN-BNL-Columbia-KEK QCD Project: T.~Blum, N.~Christ, C.~Cristian,
C.~Dawson, G.~Fleming, X.~Liao, G.~Liu, R.~Mawhinney, S.~Ohta, A.~Soni,
P.~Vranas, M.~Wingate, L.~Wu, and Y.~Zhestkov.  We thank RIKEN,
Brookhaven National Laboratory and the U.S.\ Department of Energy for
providing the facilities essential for this work.}}

\author{Tom Blum\address[RBRC]{RIKEN-BNL Research Center,
Brookhaven National Laboratory, Upton, NY 11973-5000, USA},
Shigemi Ohta\address[KEK]{Institute for Particle and Nuclear Studies, KEK,
Tsukuba, Ibaraki 305-0801, Japan}\addressmark[RBRC]
and
Shoichi Sasaki\addressmark[RBRC]\thanks{Present address: Department of
Physics, University of Tokyo, Hongo 7-3-1, Bunkyo-ku, Tokyo 113-0033,
Japan.}}
       
\begin{document}

\begin{abstract}
We present a preliminary domain-wall fermion lattice-QCD calculation of
isovector vector and axial charges, \(g_{_V}\) and \(g_{_A}\), of the
nucleon.  Since the lattice renormalizations, \(Z_{_V}\) and \(Z_{_A}\),
of the currents are identical with DWF, the lattice ratio
\((g_{_A}/g_{_V})^{\rm lattice}\) directly yields the continuum value. 
Indeed \(Z_{_V}\) determined from the matrix element of the vector
current agrees closely with \(Z_{_A}\) from a non-perturbative
renormalization study of quark bilinears. We also obtain spin related
quantities \(\Delta q/g_{_V}\) and \(\delta q / g_{_V}\).
\vspace{1pc}
\end{abstract}

\maketitle

The isovector vector and axial charges, \(g_{_V}\) and \(g_{_A}\), of the
nucleon provide an interesting additional test of the domain wall fermion
(DWF) method in the baryon sector where it has succeeded in reproducing
the mass difference between the positive- and negative-parity ground
states, \(N(939)\) and \(N^*(1535)\) \cite{nstar2000}.  These charges are
defined as
\[
g_{_V} = G_{_F} \lim_{q^2\rightarrow 0} g_{_V}(q^2)
\]
from the isovector vector current \(\langle n| V^-_\mu(x) | p \rangle\)=
\[
i\bar{u}_n [\gamma_\mu g_{_V}(q^2)
             +q_\lambda \sigma_{\lambda\mu} g_{_M}(q^2) ] u_p e^{-ipx},
\]
\[
{\rm and}\,\, g_{_A} = G_{_F} \lim_{q^2\rightarrow 0} g_{_A}(q^2)
\]
with the axial current \(\langle n| A^-_\mu(x) | p \rangle\)=
\[
i\bar{u}_n \gamma_5
             [\gamma_\mu g_{_A}(q^2)
             +q_\mu g_{_P}(q^2) ] u_p e^{-ipx}.
\]
The values of \(g_{_V} = G_{_F} \cos \theta_c\) and \(g_{_A}/g_{_V} =
1.2670(35)\) are well known from neutron \(\beta\) decay.  Here
\(G_{_F}\) denotes the Fermi constant and \(\theta_c\) the Cabibbo
angle. \(g_{_V} = G_{_F} \cos \theta_c\) follows from vector current
conservation.  In contrast the axial current should receive a strong
correction from quantum chromodynamics (QCD), resulting in the deviation
of the ratio \(g_{_A}/g_{_V}\) from unity. 

In lattice calculations in general the two relevant currents get
renormalized by the lattice cutoff.  With conventional fermion schemes
this renormalization usually makes the calculations rather difficult, if
not intractable, even for such simple quantities like \(g_{_V}\) and
\(g_{_A}\).  However with DWF it is greatly simplified because \(Z_{_A} =
Z_{_V}\)\cite{Chris}, so that the evaluation \(g_{_A}^{\rm
lattice}/g_{_V}^{\rm lattice}\) directly yields the continuum value.

Phenomenological models of baryons have not been successful in reproducing
this ratio: the non-relativistic quark model gives a value of \(5/3\),
and the MIT bag model 1.07.  Lattice calculations typically underestimate
\(g_{_A}\) by 20 \% \cite{history}.   All of these previous lattice
calculations are done with (improved) Wilson fermions  and consequently
suffer from \(Z_{_A} \neq Z_{_V}\) and other renormalization
complications.

The present numerical calculations use the same gauge configurations
reported in ref.\ \cite{zeromode}, the notations of which we follow here. 
From this work we know DWF works well. In particular: 1) fermion near-zero
mode effects are well understood,  2) small chiral symmetry breaking
induced by the finite extra dimension is described by a single parameter
\(m_{\rm res}\) in low-energy effective lagrangian, which decreases as
\(\beta\) or \(L_s\) increases, with the value of \(m_{\rm res}/m_{\rm
strange} = 0.033(3)\) at \(\beta=6.0\) and \(L_s=16\), and  3)
non-perturbative renormalization (NPR) works well for the quark bilinears
\cite{Chris}.

\begin{figure}
\includegraphics[width=70mm,bb=26 75 586 614]{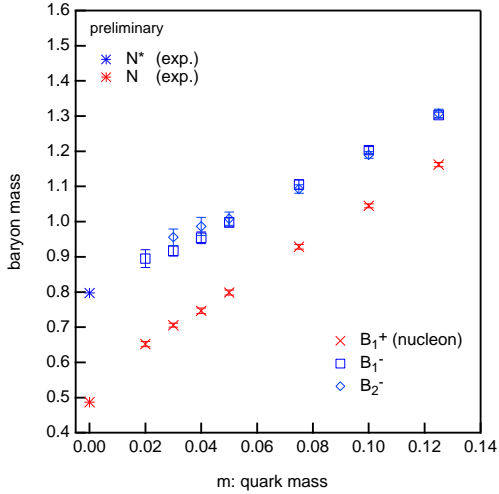}
\caption{Dependence of \(N\) (cross) and \(N^*\) (square and diamond) mass
on quark mass, \(m_f\).  Blasts at \(m_f\)=0 are experimental values in
lattice unit, \(a^{-1}\)\(\simeq\)2 GeV.}
\label{fig:nstar}
\end{figure}
Positive-parity nucleon states are created (destroyed) with interpolating
operators \(B_1^+ = \epsilon_{abc} (u_a^T C \gamma_5 d_b) u_c\) and
\(B_2^+ = \epsilon_{abc} (u_a^T C d_b) \gamma_5 u_c\) while the
negative-parity ones are created with \(B_1^- = \gamma_5 B_1^+ =
\epsilon_{abc} (u_a^T C \gamma_5 d_b) \gamma_5 u_c\) and \(B_2^- =
\gamma_5 B_2^+ = \epsilon_{abc} (u_a^T C d_b) u_c\) with appropriate
boundary conditions in time to reduce backward propagating contamination
\cite{nstar2000}.  \(B_1^+\) gives the ground-state nucleon (\(N(939)\))
mass.  On the other hand \(B_2^+\) seems to give the first excited
positive parity state for heavier bare quark mass \(m_f\).  Whether it
can reproduce the \(N'(1440)\) mass in the chiral limit is not yet
known.  \(B_1^-\) and \(B_2^-\) masses agree with each other, and yield
the negative-parity ground state, \(N^*(1535)\).
Our quenched DWF calculation reproduces very well this large mass
splitting between \(N(939)\) and \(N^*(1535)\) parity partners (see Figure
\ref{fig:nstar}).  Phenomenological models like the non-relativistic quark
model and the MIT bag model have failed here.  It should be also noted
that an earlier quenched lattice calculation using Wilson fermions
\cite{Leinweber} failed here too, though more recent calculations show
improvements \cite{LeeRichards}.

So DWF calculation of nucleon matrix elements seems promising. 
\(g_{_A}\) is interesting because it is particularly clean with DWF since
\(Z_{_A}\)=\(Z_{_V}\).  It is also interesting to see how well quenched
calculations work for a well-known example of soft-pion behavior, namely
the Goldberger-Treiman relation:
\(g_{_A}/g_{_V}\)\(\simeq\)\(f_\pi g_{\pi{_N}{_N}} /m_{_N}\)\(\simeq
1.31\).  We know that with DWF the ratio \(f_\pi / m_{_N}\) is almost
constant over the range of \(m_f\) we are using, and agrees well with
the experimental value \cite{zeromode}.

We follow the standard practice \cite{history} for our two- and
three-point function calculations.  The two-point function is defined by
\[
G_{_{_N}}(t) = {\rm Tr}[
(1+\gamma_t) \sum_{\vec{x}} \langle TB_1(x)B_1(0) \rangle ],
\]
using \(B_1 = \epsilon_{abc} (u_a^T C \gamma_5d_b)u_c\) for the proton. 
The three-point function for the local vector current is
\(G_{_V}^{u,d}(t,t')\),
\[
{\rm Tr} [
(1+\gamma_t) \sum_{\vec{x}', \vec{x}}
\langle TB_1(x) V_t^{u,d}(x') B_1(0) \rangle ],
\]
and for the local axial current, \(G_{_A}^{u,d}(t,t')\),
\[
{\rm Tr} [
(1+\gamma_t)\gamma_i\gamma_5 \sum_{\vec{x}', \vec{x}}
\langle TB_1(x) A_i^{u,d}(x') B_1(0) \rangle ],
\]
averaged over \(i=x, y, z\).  We choose a fixed \(t=t_{\rm
source}-t_{\rm sink}\) and \(t' < t\).  From their lattice estimates
\[
g_{_\Gamma}^{\rm lattice}
= \frac{G_{_\Gamma}^u(t,t')-G_{_\Gamma}^d(t,t')}{G_{_{_N}}(t)},
\] 
with \(\Gamma = V\) or \(A\), the continuum values
\[
g_{_\Gamma} = Z_{_\Gamma} g_{_\Gamma}^{\rm lattice},
\]
are obtained.  Here we need the non-perturbative renormalizations, defined
by
\[
[\bar{u} \Gamma d]_{\rm ren} = Z_{_\Gamma} [\bar{u}\Gamma d]_0,
\]
\begin{figure}
\includegraphics[width=70mm,bb=18 69 578 593]{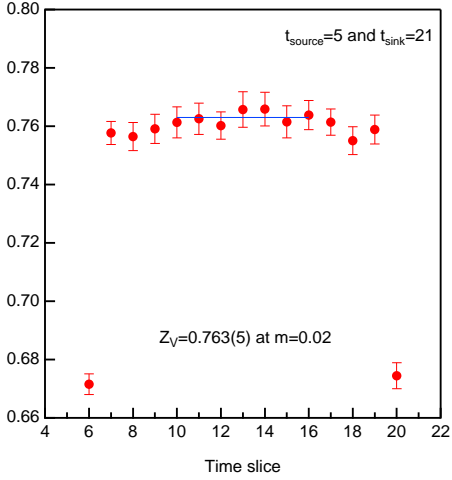}
\caption{Dependence of vector renormalization, \(Z_{_V} = 1/g_{_V}^{\rm
lattice}\), on \(t'\), at \(m_f=0.02\).  A good plateau is observed.}
\label{fig:zv}
\end{figure}
which should satisfy \(Z_{_A} = Z_{_V}\) so that
\[
\left(\frac{g_{_A}}{g_{_V}}\right)^{\rm continuum} =
\left(
\frac{G_{_A}^u(t,t')-G_{_A}^d(t,t')}{G_{_V}^u(t,t')-G_{_V}^d(t,t')}
\right)^{\rm lattice}.
\]
Note \(g_{_A}\) is described as \(\Delta u - \Delta d\) where \(\Delta
f\) (\(f=u\) or \(d\)) is defined by
\[
\langle p,s | \bar{f} \gamma_5 \gamma_\mu
f | p,s \rangle = 2 s_\mu \Delta q,
\]
with \(s\) satisfying \(s\cdot p = 0\) and \(s^2 = -1\).  From these we
obtain spin-polarized longitudinal parton distribution, \(\Delta q = \int
dx [q_{\uparrow}(x) - q_{\downarrow}(x)] = \Delta u + \Delta d\). 
Similarly, \(\delta f\) is defined by
\[
\langle p,s | i \bar{f} \sigma_{\mu\nu}\gamma_5 f | p,s \rangle
= 2 (s_\mu p_\nu - s_\nu p_\mu) \delta f,
\]
with \(\sigma_{\mu\nu} = [\gamma_\mu, \gamma_\nu]/2\).  This gives
the tensor charge which is related to the transverse parton distribution,
\(\delta q = \int dx [q_{\bot}(x) - q_{\top}(x)] = \delta u + \delta d\). 
We define \(G_{_T}^q(t,t')\) by inserting \(T_i^q = \bar{q}\gamma_t
\gamma_i \gamma_5 q\) at \(t'\) and a projection operator 
\((1+\gamma_{t})\gamma_{i}\gamma_{5}\), and
\[
\delta q ^{\rm lattice} = \frac{G_{_T}^u(t,t') +
G_{_T}^d(t,t')}{G_{_N}(t)}
\]
is obtained.  Here we need \(Z_{_T}\), which is scheme- and
scale-dependent.  Note that \(\Delta u = \delta u = 4/3\) and \(\Delta d =
\delta d = -1/3\) in the heavy quark limit.

\begin{figure}[t]
\includegraphics[width=70mm,bb=28 65 588 589]{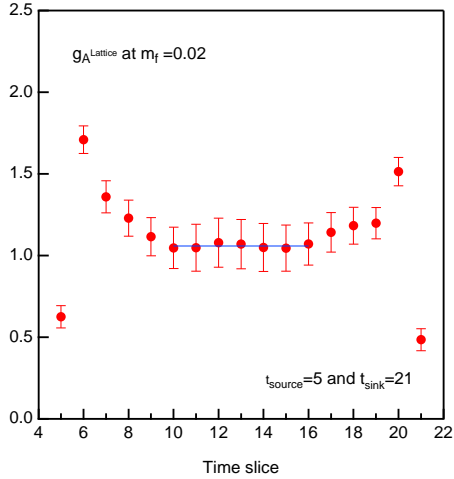}
\caption{The lattice axial charge, \(g_{_A}^{\rm lattice}\), at
\(m_f=0.02\).  A good plateau is seen in \(10 \le t \le 16\).}
\label{fig:za}
\end{figure}

The numerical calculations are from 200 configurations at \(\beta=6.0\) on
a \(16^3 \times 32\) lattice with  DWF parameters \(L_s=16\) and
\(M_5=1.8\).   We set the source at \(t=5\), sink at 21, and current
insertions in between.  The vector renormalization,
\(Z_{_V} = 1/g_{_V}^{\rm lattice}\), is well-behaved,
\begin{figure}[t]
\includegraphics[width=70mm,bb=26 93 586 617]{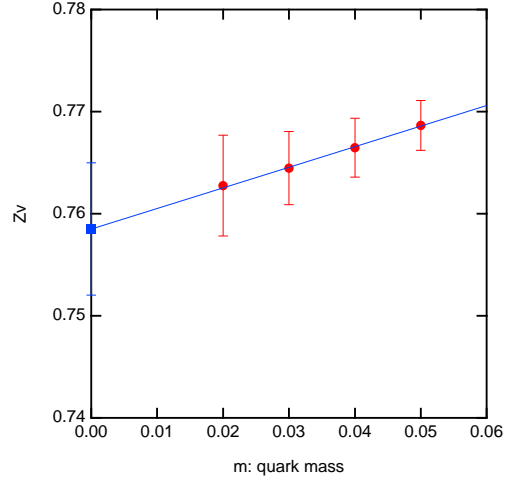}
\caption{Dependence of vector renormalization, \(Z_{_V} = 1/g_{_V}^{\rm
lattice}\), on \(m_f\).  Note the scale.  Slight linear dependence
extrapolates to the value of 0.759(6) at \(m_f\)=0.}

\label{fig:zv_ave}
\end{figure}
The value \(0.763(5)\) at \(m_f = 0.02\) (See Figure \ref{fig:zv}) agrees
well with \(Z_{_A} = 0.7555(3)\), obtained from \(\langle A^{\rm
conserved}_\mu (t) \bar{q} \gamma_5 q (0) \rangle = Z_{_A} \langle A^{\rm
local}_\mu (t) \bar{q} \gamma_5 q (0) \rangle\)  \cite{zeromode}.
A linear extrapolation gives \(Z_{_V} = 0.759(6)\) at \(m_f = 0\) (Figure
\ref{fig:zv_ave}).  For the lattice axial charge, \(g_{_A}^{\rm
lattice}\), plateaus are seen for \(10 \le t \le 16\), with a fairly strong
dependence on \(m_f\) (See for example Figure \ref{fig:za}).  So the
charge ratio, \(g_{_A}/g_{_V}\), averaged in \(10\le t \le 16\),
\begin{figure}[t]
\includegraphics[width=70mm,bb=26 77 586 601]{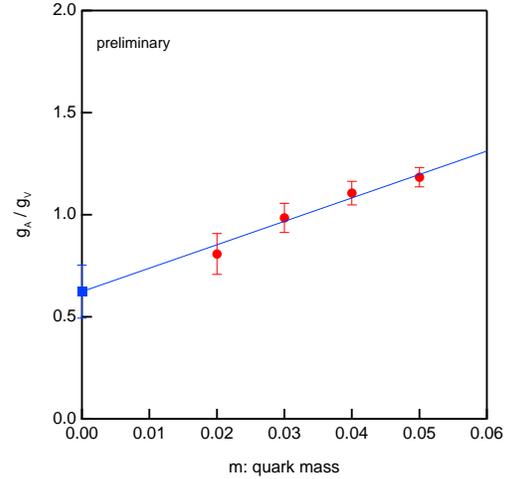}
\caption{Dependence of \(g_{_A}/g_{_V}\) on \(m_f\).}
\label{fig:gagv}
\end{figure}
linearly extrapolates to 0.62(13) at \(m_f=0\) (Figure \ref{fig:gagv})
which is about a factor of 2 smaller than experiment.  However a linear
fit may not be justified here. There is some curvature apparent in Figure
\ref{fig:gagv}, so the  value of \(g_{_A}/g_{_V}\) in the chiral limit
may be even lower. The same calculation yields (with linear extrapolations
to \(m_f=0\)) \(\Delta u / g_{_V} = 0.50(12)\) and \(\Delta d / g_{_V} =
-0.14(6)\).  Similarly, \(\delta u / g_{_V} = 0.39(11)\) and \(\delta d /
g_{_V} = -0.11(4)\).  A preliminary value for \(Z_{_T}/Z_{_A}\) is 1.1(1)
\cite{Chris}.

In summary we have explored the isovector weak interaction of the nucleon
in lattice QCD with domain-wall fermions.  All the relevant
three-point functions are well behaved.  \(Z_{_V}\) determined from the
matrix element of the vector current agrees closely with that from an NPR
study of quark bilinears \cite{Chris}.  Linear extrapolations to
\(m_f\)=0 give
\begin{itemize}
\item \(g_{_A}/g_{_V} = 0.62(13)\),
\item \(\Delta q /g_{_V} = 0.36(14)\),
\item \(\delta q / g_{_V} = 0.31(12)\).
\end{itemize}
The quite low value of \(g_{_A}/g_{_V}\) that we obtained requires further
investigation. In particular, we are studying the Ward-Takahashi identity
which governs \(g_{_A}\).  If the matrix element of the pseudoscalar
density does not develop a pole as \(m_f\to 0\) which is expected in the
Goldberger-Treiman relation, the left hand side, and therefore \(g_{_A}\),
must vanish. Further study is also required to check systematic errors
arising from finite lattice volume, excited states (small separation
between \(t_{\rm source}\) and \(t_{\rm sink}\)), and quenching (zero
modes, absent pion cloud, etc), especially in the lighter quark mass
region.

\end{document}